\documentclass[prl,twocolumn,showpacs,preprintnumbers,amsmath,amssymb]{revtex4}

\usepackage{epsfig}
\usepackage{graphicx}
\usepackage{dcolumn}
\usepackage{bm}

\begin{document}


\title{Correlation-function asymptotic expansions: pre-factors universality}
\author{J. M. P. Carmelo$^{1,2}$ and K. Penc$^3$} \affiliation{$^1$Department of Physics,
Massachusetts Institute of Technology, Cambridge, MA
02139-4307}\affiliation{$^2$GCEP - Center of Physics, University
of Minho, P-4710-057 Braga, Portugal}\affiliation{$^3$Research
Institute for Solid State Physics and Optics, H-1525 Budapest,
P.O.B. 49, Hungary}
\date{29 August 2005}


\begin{abstract}
We show that the pre-factors of all terms of the one-dimensional
Hubbard model correlation-function asymptotic expansions have an
universal form, as the corresponding critical exponents. In
addition to calculating such pre-factors, our study clarifies the
relation of the low-energy Tomonaga-Luttinger liquid behavior to
the scattering mechanisms which control the spectral properties of
the model at all energy scales. Our results are of general nature
for many integrable interacting models and provide a broader
understanding of the unusual properties of quasi-1D
nanostructures, organic conductors, and optical lattices of
fermionic atoms.
\end{abstract}


\maketitle

Over the past 25 years it has been found that the low-energy
physics of a variety of models of one-dimensional (1D) correlated
electrons can be described by the Tomonaga-Luttinger liquid (TLL)
theory \cite{Voit}. Importantly, the low-energy TLL universal
behavior was observed in different real materials and systems, as
for instance in carbon nanotubes \cite{Hiro}, ballistic wires
\cite{Halperin}, quasi-1D organic conductors \cite{Lorenz}, and
quasi-1D quantum gases of ultracold fermionic atoms \cite{Recati}.
On the other hand, the low-energy phases of some quasi-1D
compounds are not metallic and correspond to broken-symmetry
states \cite{spectral0}. Recently, the resolution of photoemission
experiments has improved, and the {\it normal} state of these
compounds was found to display exotic spectral properties
\cite{spectral0}. However, such a metallic phase refers to finite
energies and is not described by the TLL theory.

The 1D Hubbard model is one of the few realistic models for
correlated electrons in a discrete lattice for which one can
exactly calculate all the energy eigenstates and their energies
\cite{Lieb}. It includes a first-neighbor transfer-integral for
electron hopping along the chain and an effective on-site
repulsion $U$. For finite energy values, the metallic phase of
this model is beyond a TLL description and thus the study of
spectral functions is a very involved many-electron problem.
Fortunately, the recently introduced {\it pseudofermion dynamical
theory} (PDT) provides explicit expressions for these functions
\cite{V-1}. Moreover, the theory describes successfully the
unusual spectral features of quasi-1D compounds for the whole
finite-energy band width \cite{spectral}. Recently, consistent
results were obtained by numerical techniques \cite{Eric}.
Furthermore, when combined with the Renormalization Group, the use
of the PDT reveals that a system of weakly coupled Hubbard chains
is suitable for the successful description of the phase diagram
observed in quasi-1D doped Mott-Hubbard insulators \cite{super}.
In turn, the low-energy physics of the model corresponds to the
universal TLL behavior and was studied by different techniques,
such as bosonization \cite{Schulz} and conformal-field theory
(CFT) \cite{Belavin}.

There are many investigations where the low-energy conformal invariance was combined with
the model exact Bethe-ansatz solution in the study of the asymptotics of correlation
functions and related quantities \cite{Woy,Ogata,Kawakami,Frahm,Brech,Karlo}. However,
the relation of the low-energy TLL behavior to the microscopic scattering mechanisms
which control the unusual spectral properties of the model at all energy scales remains
an interesting open problem, which we address in this paper. While CFT and bosonization
techniques do not provide correlation-function expressions for finite energy, here we use
the general PDT to derive the asymptotic expansions for the model correlation functions.
Such asymptotic expansions go beyond those provided by other methods, once we can derive
explicit expressions for the pre-factors of all the expansion terms. Moreover, we find
that such pre-factors have an universal form, as the corresponding critical exponents.
Our results describe the emergence of the TLL low-energy physics in terms of the general
non-perturbative microscopic scattering mechanisms of the model at all energy scales and
lead to a broader understanding of the unusual properties observed in low-dimensional
materials and nanostructures \cite{Hiro,Halperin,Lorenz} and systems of interacting
ultracold fermionic atoms in 1D optical lattices \cite{Jaksch}.

In our study we consider the 1D Hubbard model with periodic
boundary conditions, number of lattice sites $N_a$ large and even,
and units such that the transfer integral and the Planck and
electronic-lattice constants are one. Thus, the lattice length is
$L=N_a$. Let us consider the general correlation function,
\begin{equation}
\chi_{{\cal{N}}}^{l} (k,\omega) =
l\int_{-\infty}^{+\infty}d\omega'\,{B_{{\cal{N}}}^{l}
(k,\omega')\over\omega -\omega' +il0} \, , \label{chi0}
\end{equation}
where $B_{{\cal{N}}}^{l} (k,\omega)$ is the $\cal{N}$-electron
spectral function,
\begin{equation}
B_{{\cal{N}}}^{l} (k,\omega) = \sum_{f}\, \vert\langle f\vert\,
{\hat{O}}_{{\cal{N}}}^{l} (k) \vert GS\rangle\vert^2\,\delta
(\omega - l\Delta E_f) \, . \label{ABON}
\end{equation}
Here the index $l$ can have the values $l = \pm 1$, $l\omega
> 0$, the general $\cal{N}$-electron operators ${\hat{O}}_{{\cal{N}}}^{+1} (k)\equiv
{\hat{O}}_{{\cal{N}}}^{\dag} (k)$ and ${\hat{O}}_{{\cal{N}}}^{-1}
(k) \equiv {\hat{O}}_{{\cal{N}}} (k)$ carry momentum $k$, the $f$
summation runs over the excited energy eigenstates, and $\Delta
E_f=[E_f - E_{GS}]$ where the energy $E_f$ corresponds to these
final states and $E_{GS}$ is the ground-state energy. Most common
examples are the operator ${\hat{O}}_{1} (k) = c_{k,\sigma}$ and
different choices of charge, spin, and Cooper-pair ${\cal{N}}=2$
operators. We consider that the electronic density $n=n_{\uparrow
}+n_{\downarrow}$ and spin density $m=n_{\uparrow}-n_{\downarrow}$
are in the range $0<n<1$ and $0<m<n$ and thus the Fermi momenta
are given by $k_F=\pi n/2$ and $k_{F\sigma}=\pi n_{\sigma}$.

The Fourier transform ${\tilde{\chi}}_{{\cal{N}}}^{l} (x,t)$ of
the function (\ref{chi0}) relative to both the momentum $k$ and
energy $\omega$ can be expressed in terms of the corresponding
Fourier transform $\tilde{B}_{{\cal{N}}}^{l} (x,t)$ of the
spectral function (\ref{ABON}) as follows,
\begin{equation}
{\tilde{\chi}}_{{\cal{N}}}^{l} (x,t) = -i2\pi\,\theta
(lt)\,\tilde{B}_{{\cal{N}}}^{l} (x,t) \, , \label{FTchi0}
\end{equation}
where here and in other expressions of this paper $\theta (y)=1$
for $y>0$ and $\theta (y)=0$ for $y\leq 0$.

Our main goal is the study of the general correlation function
(\ref{FTchi0}) asymptotic expansion. Such an asymptotic expansion
is controlled by the low-energy behavior of the function
(\ref{ABON}). From the studies of Ref. \cite{I} one knows that the
excitation spectrum is such that for the densities considered
here, the excited energy eigenstates which span the low-energy
subspace can be represented by occupancy configurations of the $c$
and $s1$ pseudoparticles, which are defined in terms of the
electrons in that reference. We denote the $s1$ branch by $s$ and
use a general index $\alpha$ such that $\alpha =c,s$. The $\alpha$
pseudoparticles have discrete bare-momentum values $q_j$ such that
$q_{j+1}-q_j=2\pi/L$ and energy residual-interaction terms
associated with $f$ functions $f_{\alpha ,\alpha'} (q,q')$. In the
continuum limit the bare-momentum values $q$ exist in the range
$\vert q\vert\leq q^0_{\alpha}$ where except for $1/L$ corrections
$q^0_c$ and $q^0_s$ read $q^0_c =\pi$ and $q^0_s=k_{F\uparrow}$
and play the role of {\it Brillouin zone} limits. It is useful to
consider the index $\iota =\pm 1$ associated with the right
($\iota =+1$) and left ($\iota =-1$) $\alpha,\iota$ {\it Fermi
points}, which except for $1/L$ corrections are given by $\iota
q^0_{Fc} = \iota 2k_F$ and $\iota q^0_{Fs} = \iota
k_{F\downarrow}$ for the ground state. That state corresponds to
$\alpha$-band bare-momentum densely packed occupancy
configurations such that $\vert q\vert\leq q^0_{F\alpha}$ and
$q^0_{F\alpha}<\vert q\vert \leq q^0_{\alpha}$ for $\alpha$
pseudofermions and $\alpha$ pseudofermion holes, respectively.
(All $\alpha$ and $\iota$ sums or products appearing in the
expressions provided below run over the values $\alpha =c,s$ and
$\iota =+1,-1$, respectively.)

The pseudoparticle - pseudofermion unitary transformation plays a
key role in the PDT. For the low-energy subspace, such a
transformation maps the $\alpha$ pseudoparticles or pseudoparticle
holes onto $\alpha$ pseudofermions or pseudofermion holes, by
introducing shifts, $Q^{\Phi}_{\alpha} (q_j)/L$, of order $1/L$ in
the discrete bare-momentum values and leaving all other
pseudoparticle properties invariant \cite{V-1,S-P}. The $\alpha$
pseudofermions have discrete {\it canonical-momentum values}
${\bar{q}}_j = q_j + Q^{\Phi}_{\alpha} (q_j)/L$ and no energy
residual-interaction terms. The point is that while for the
electrons the present problem is strongly correlated, for the
pseudofermions the shift $Q^{\Phi}_{\alpha} (q)/L$ results from
zero-momentum forward scattering only. For the ground state,
${\bar{q}}_j = q_j$ and the pseudoparticle and pseudofermion are
the same object. Each ground-state - excited-state transition
involves a scattering event whose scattering centers are the
$\alpha',q'$ pseudofermions and pseudofermion holes created under
the transition and the scatterers are all $\alpha,q$
pseudofermions and pseudofermion holes of the excited state, which
acquire an overall scattering phase shift $Q^{\Phi}_{\alpha}
(q)/2=
\sum_{\alpha'}\sum_{q'}\pi\,\Phi_{\alpha\,\alpha'}(q,q')\,\Delta
N_{\alpha'} (q')$. Here $\Delta N_{\alpha'} (q')$ is the
excited-state $\alpha'$-band bare-momentum distribution function
deviation and $\pm \pi\,\Phi_{\alpha\,\alpha'}(q,q')$ is a
two-pseudofermion phase shift acquired by the $\alpha,q$
pseudofermion or hole scatterers for each $\alpha',q'$
pseudofermion $(+)$ or hole $(-)$ scattering center created under
the transition. The value of such a phase shift is in the range
$-\pi/2\leq\pi\,\Phi_{\alpha\,\alpha'}(q,q')\leq+\pi/2$ and its
$U$, $n$, and $m$ dependence is uniquely defined by solution of a
system of integral equations. The overall phase shift reads,
\begin{equation}
Q_{\alpha}(q)/2 = Q_{\alpha}^0/2 + Q^{\Phi}_{\alpha} (q)/2 \, ,
\label{Qcan1j}
\end{equation}
where $Q_{\alpha}^0/2=0,\pm\pi/2$ is a scattering-less
contribution which has a single and well-defined value for the
whole low-energy excitation subspace of each correlation function.
Such a subspace is spanned by all energy eigenstates with the same
value for the number deviations $\Delta N_{c}=\Delta N$ and
$\Delta N_{s}=\Delta N_{\downarrow}$ such that $\Delta
N_{\alpha'}=\sum_{q'}\Delta N_{\alpha'} (q')= \sum_{\iota}\Delta
N_{\alpha',\iota}^F$. In contrast to
$\pi\,\Phi_{\alpha\,\alpha'}(q,q')$, $Q^{\Phi}_{\alpha} (q)/2$ is
a functional whose value depends on the excited state trough the
deviations $\Delta N_{\alpha'} (q')$. For the present low-energy
problem, such deviations refer to $q'$ values in the vicinity of
the {\it Fermi points} only and thus the number $\Delta
N_{\alpha',\iota}^F = \Delta N^{0,F}_{\alpha',\iota}+\iota\,Q^
0_{\alpha'}/2\pi $ is such that $\Delta N^{0,F}_{\alpha',\iota}$
is the deviation in the number of $\alpha'$ pseudofermions at the
$\alpha',\iota$ {\it Fermi point}. The $\alpha'$-branch
current-number deviation reads $\Delta J_{\alpha'}^F ={1\over
2}\sum_{\iota}\iota\Delta N_{\alpha',\iota}^F$. The low-energy
subspace contains several {\it J-subspaces}, which differ at least
in one of the two values $\{\Delta J_{c}^F,\Delta J_{s}^F\}$. The
{\it J-ground state} is the lowest-energy eigenstate of a
J-subspace. Its bare-momentum occupancy configurations are densely
packed with {\it Fermi points} $q_{F\alpha,\iota} =
\iota\,q^0_{F\alpha} + \Delta q_{F\alpha,\iota}$. The deviation
$\Delta q_{F\alpha,\iota}$ and the excitation momentum of that
state read,
\begin{equation}
\Delta q_{F\alpha,\iota}  = \iota {2\pi\over L} \Bigl({\Delta
N_{\alpha}\over 2} +\iota\Delta J^F_{\alpha}\Bigr) \, ;
\hspace{0.25cm} k^F_0 = \sum_{\alpha'} 2q^0_{F\alpha'}\Delta
J_{\alpha'}^F \, . \label{qF}
\end{equation}
The momentum $k^F_0$ is generated by zero-energy and
finite-momentum elementary processes, which create $\alpha'$
pseudofermions ($\Delta N^{0,F}_{\alpha',\iota}>0$) or $\alpha'$
pseudofermion holes ($\Delta N^{0,F}_{\alpha',\iota}<0$) at least
at one of the four $\alpha',\iota$ {\it Fermi points}. The
corresponding J-subspace is spanned by energy eigenstates
generated from the J-ground state by small-momentum and low-energy
"particle-hole" pseudofermion processes in the vicinity of the
$\alpha',\iota$ {\it Fermi points}. Such processes conserve the
set of $\{\Delta N_{c},\Delta N_{s},\Delta J_{c}^F,\Delta
J_{s}^F\}$ deviation values. A crucial point for the low-energy
scattering properties is that the $\alpha'$ pseudofermions and
holes created by such small-momentum and low-energy processes are
not active scattering centers, once the phase shifts generated by
the created pseudofermions exactly cancel those originated by
creation of the corresponding holes. It follows that the overall
phase shift (\ref{Qcan1j}) has for each $\alpha$ pseudofermion or
hole scatterer of bare-momentum $q$ the {\it the same value} for
all excited states spanning the same J-subspace with,
\begin{equation}
Q^{\Phi}_{\alpha} (q)/2 = \pi
\sum_{\alpha'}\sum_{\iota'}\Phi_{\alpha\,\alpha'}(q,\iota'q^0_{F\alpha'})
\Bigl({\Delta N_{\alpha'}\over 2} +\iota'\Delta
J^F_{\alpha'}\Bigr) \, . \label{qcan1j}
\end{equation}

A mechanism which plays a central role in the spectral properties
at all energy scales is the occurrence of a shift in the value of
the four $\alpha,\iota$ canonical-momentum {\it Fermi-points}, as
a result of each ground-state - excited-J-ground-state transition.
The square of that shift reads,
\begin{eqnarray}
2\Delta_{\alpha}^{\iota} & \equiv & \Bigl(\Delta
{\bar{q}}_{F\alpha,\iota}/ {2\pi\over L}\Bigr)^2 \nonumber \\
& = & \Bigl(\iota\,\Delta N_{\alpha,\iota}^{0,F}+ {Q_{\alpha}
(\iota\,q^0_{F\alpha})\over 2\pi}\Bigr)^2 \, . \label{Delta}
\end{eqnarray}
At low energy, consistently with the form of the scattering phase
shift (\ref{qcan1j}), $2\Delta_{\alpha}^{\iota}$ simplifies to,
\begin{eqnarray}
2\Delta_{\alpha}^{\iota} & = & 2\Delta_{\alpha}^{\iota} (\Delta
N_c,\Delta N_s,\Delta J_c^F,\Delta J_s^F) \nonumber \\
& = &
\Bigl(\sum_{\alpha'}\Bigl[\iota\,\xi^0_{\alpha\,\alpha'}\,{\Delta
N_{\alpha'}\over 2} + \xi^1_{\alpha\,\alpha'}\,\Delta
J^F_{\alpha'}\Bigr]\Bigr)^2 \, . \label{Delta0}
\end{eqnarray}
Here $\xi^j_{\alpha\,\alpha'}= \delta_{\alpha,\alpha'} +
\sum_{\iota=\pm 1}(\iota^j)\,\Phi_{\alpha\,\alpha'}
(q^0_{F\alpha},\iota\,q^0_{F\alpha'})$ with $j=0,1$ involves two
two-pseudofermion phase shifts whose value is such that in the
present low-energy limit the square of the shifts in the value of
the $\alpha,\iota$ pseudofermion {\it Fermi-points} given in Eqs.
(\ref{Delta}) and (\ref{Delta0}) equals the conformal dimension of
the corresponding CFT $\alpha,\iota$ primary fields. Within the
PDT the important functional (\ref{Delta}) is well defined for all
energy scales and corresponds to a much more general paradigm
\cite{V-1}. Thus, this connection only emerges in the low-energy
limit considered here.

The evaluation of the asymptotic expansion of the function
$\tilde{B}_{{\cal{N}}}^{l} (x,t)$ on the right-hand side of Eq.
(\ref{FTchi0}) involves the use of the general PDT expressions for
the corresponding spectral function $B_{{\cal{N}}}^{l} (k,\omega)$
of Eq. (\ref{ABON}). After some algebra, we find the following
general asymptotic expansion for the correlation function
(\ref{FTchi0}),
\begin{eqnarray}
{\tilde{\chi}}_{{\cal{N}}}^{l} (x,t) & = & \theta
(lt)\sum_{\{\Delta
J^F_{\alpha}\}}\Bigl\{e^{ilk^F_0\,x}\,\chi_0\nonumber \\
& \times & \prod_{\alpha}\prod_{\iota} \Bigl({1\over
x-\iota\,v_{\alpha}\,t+i\iota\,0}\Bigr)^{2\Delta_{\alpha}^{\iota}}\Bigr\}\,
, \label{tiCHI}
\end{eqnarray}
where $v_{\alpha}$ is the $\alpha$ pseudofermion group velocity at
$q^0_{F\alpha}$ and each term of the $\sum_{\{\Delta
J^F_{\alpha}\}}$ summation is generated by transitions from the
ground state to the excited states with the same values for the
current deviations $\{\Delta J^F_{c},\Delta J^F_{s}\}$. Note that
the asymptotic expansion (\ref{tiCHI}) has the expected general
form, which coincides with that provided by CFT and used in the
studies of Refs. \cite{Woy,Ogata,Kawakami,Frahm,Brech,Karlo}.
However, here we could obtain an explicit expression for the
pre-factors $\chi_0 =\chi_0 (\Delta N_c,\Delta N_s,\Delta
J_c^F,\Delta J_s^F)$ of such an asymptotic expansion. It is given
by,
\begin{equation}
\chi_0 = -i{\pi\over 2}\,e^{-i{\pi\over 2}\lambda_l}\prod_{\alpha} \Lambda_{\alpha} \, ,
\label{chi00}
\end{equation}
where
\begin{equation}
\lambda_l = l\sum_{\alpha} \sum_{\iota}\iota
2\Delta_{\alpha}^{\iota} \, ; \hspace{0.35cm} \Lambda_{\alpha} =
\left({N_a\over 2\pi}\right)^{\sum_{\iota}
2\Delta_{\alpha}^{\iota}} {A^{(0,0)}_{\alpha}\over \sqrt{N_a}} \,
, \label{Lambda}
\end{equation}
and the $\alpha$ pseudofermion weight reads,
\begin{widetext}
\begin{eqnarray}
& &  A^{(0,0)}_{\alpha} = A^{(0,0)}_{\alpha} (\Delta N_c,\Delta
N_s,\Delta J_c^F,\Delta J_s^F) = \Big({1\over
N_{\alpha}^*}\Bigr)^{2[N^0_{\alpha}+\Delta
N_{\alpha}]}\prod_{q_j\in {\cal F}}\sin^2
\frac{Q_{\alpha}(q_j)}{2}\prod_{j=1}^{N_{\alpha}^*-1}
\left(\sin \frac{\pi j}{N_{\alpha}^*}\right)^{2(N_{\alpha}^* -j)}\nonumber \\
& \times & \prod_{q_i\in {\cal F}} \prod_{q_j\in {\cal F}}\theta
(q_j-q_i)\,\sin^2\frac{Q_{\alpha}({q}_j)/2 - Q_{\alpha}({q}_i)/2 +
\pi (j-i)}{ N_{\alpha}^*}\prod_{q_i\in {\cal F}} \prod_{q_j\in
{\cal F}} \sin^{-2}{\pi (j-i)+Q_{\alpha}({q}_j)/2\over
N_{\alpha}^*}  \, . \label{A00}
\end{eqnarray}
\end{widetext}
Here $N_{c}^* = N_a$, $N_{s}^* = N^0_{\uparrow}+\Delta
N_{\uparrow}$, $Q_{\alpha}(q_j)/2$ is the overall phase shift
(\ref{Qcan1j}), and $q_j\in {\cal F}$ corresponds to the set of
discrete bare-momentum values in the range $q_{F\alpha,-1}\leq
q_j\leq q_{F\alpha,+1}$ where $q_{F\alpha,\iota} =
\iota\,q^0_{F\alpha} + \Delta q_{F\alpha,\iota}$ is the J-ground
state bare-momentum {\it Fermi point} whose deviation is given in
Eq. (\ref{qF}). The $N_a$ dependence of $A^{(0,0)}_{\alpha}$ is
such that the quantity $\Lambda_{\alpha}$ given in Eq.
(\ref{Lambda}) is independent of $N_a$.

Our study refers to the universal part of the asymptotic expansion
of correlation functions, Eq. (\ref{tiCHI}), and does not involve
the logarithmic corrections, which are specific to each
correlation function \cite{Schulz}. The universal character of the
asymptotic expansion (\ref{tiCHI}) is such that the value of the
conformal dimensions only depends on the specific correlation
function through the values of the four deviations $\Delta N_c$,
$\Delta N_s$, $\Delta J_c^F$, and $\Delta J_s^F$ of each allowed
excitation J-subspace. Otherwise, the $U$, $n$, and $m$ dependence
of the two-pseudofermion phase-shift parameters
$\xi^j_{\alpha\,\alpha'}$ of Eq. (\ref{Delta0}) is specific to the
model but is the same for all its correlation functions.
Importantly, Eqs. (\ref{chi00})-(\ref{A00}) reveal that the same
occurs with the pre-factors $\chi_0$ of the asymptotic expansion
(\ref{tiCHI}). Indeed, the expression of the associated weight of
Eq. (\ref{A00}) involves the overall phase shift $Q_{\alpha}(q)/2$
given in Eqs. (\ref{Qcan1j}) and (\ref{qcan1j}), which for each
value of $q$ also depends on the specific correlation function
through the values of the four deviation numbers $\Delta N_c$,
$\Delta N_s$, $\Delta J_c^F$, and $\Delta J_s^F$ of each allowed
J-subspace. Moreover, the products of the $A^{(0,0)}_{\alpha}$
expression run in the range $q_{F\alpha,-1}\leq q_j\leq
q_{F\alpha,+1}$, whose limiting values deviations are given in Eq.
(\ref{qF}) and are solely determined by the deviations $\Delta
N_{\alpha}$ and $\Delta J_{\alpha}^F$. Thus, the pre-factors
$\chi_0 = \chi_0 (\Delta N_c,\Delta N_s,\Delta J_c^F,\Delta
J_s^F)$ value also depends on the specific correlation function
through the values of the four deviation numbers $\Delta N_c$,
$\Delta N_s$, $\Delta J_c^F$, and $\Delta J_s^F$ only, as the
conformal dimensions. Otherwise, the $U$, $n$, and $m$ dependence
of the two-pseudofermion phase shifts involved in the $\chi_0$
expression is specific to the model but is again the same for all
its correlation functions.

In this paper we obtained the pre-factor $\chi_0$ of each term of the asymptotic
expansion (\ref{tiCHI}) for the correlation functions of the 1D Hubbard model. The form
of these pre-factors, Eq. (\ref{chi00}), is universal for all correlation functions.
Their value is controlled by the overall pseudofermion and hole phase shifts, Eq.
(\ref{Qcan1j}), through the dependence on these shifts of the two weights
$A^{(0,0)}_{\alpha}$ of Eq. (\ref{A00}) and four functionals $2\Delta_{\alpha}^{\iota}$
of Eqs. (\ref{Delta}) and (\ref{Delta0}). Concerning the relation of the latter
quantities to the scattering mechanisms, note that in the $A^{(0,0)}_{\alpha}$ expression
(\ref{A00}) the bare-momentum products run over the overall phase shifts of the $\alpha$
pseudofermion scatterers with bare momentum inside the J-ground-state {\it Fermi sea},
whose scattering centers are the $c$ and $s$ pseudofermion and holes created at the
J-ground-state {\it Fermi points}. Furthermore, the four conformal dimensions of the CFT
primary fields equal the square of the shifts in the two $c$ and two $s$ pseudofermion
canonical-momentum {\it Fermi points}. The four functionals $2\Delta_{\alpha}^{\iota}$
and the two weights $A^{(0,0)}_{\alpha}$ also play an important role in the finite-energy
scattering properties, by controlling the unusual spectral properties of the model
\cite{V-1,S-P} and real materials \cite{spectral,super} at all energy scales. Thus, our
results reveal the connection of the low-energy quantities to the scattering mechanisms
that control the spectral properties at all energy scales.

While in this paper we considered the 1D Hubbard model, our results are of general nature
for many integrable interacting problems and therefore have wide applicability. Such
results provide a broader understanding of existing quasi-1D systems and materials
\cite{Hiro,Halperin,Lorenz,Jaksch}. (Recently, the model was used in preliminary
investigations of the density profiles of 1D ultracold fermionic atoms in an optical
lattice \cite{Liu}.) Indeed, our results relate their low-energy properties to the the
general scattering processes of the objects whose occupancy configurations describe the
exotic quantum phases of matter corresponding to their different energy scales. This is
confirmed for finite energies in Refs. \cite{spectral0,spectral}, where the general PDT
weight distributions are shown to describe the photoemission features of quasi-1D
compounds for the whole finite-energy band width, whereas the TLL universal behavior was
observed in quasi-1D materials and systems whose low-energy phase is metallic
\cite{Hiro,Halperin,Lorenz,Recati}.

We thank Y. Chen, P. A. Lee, S. -S. Lee, and T. C. Ribeiro for
discussions and the support of MIT, Gulbenkian Foundation,
Fulbright Commission, ESF Science Program INSTANS 2005-2010, OTKA
grant T049607, and FCT grant POCTI/FIS/58133/2004.

\end{document}